\documentclass[pra,twocolumn]{revtex4}
\usepackage[dvips]{graphicx}%
\usepackage{bm,color}
\usepackage{amsmath,amssymb}

\newcommand{\ketbra}[2]{\ket{#1}\bra{#2}} 
\newcommand{\ket}[1]{\left |  #1 \right \rangle}
\newcommand{\bra}[1]{ \left \langle #1  \right |}
\newcommand{\ave}[1]{  \langle #1   \rangle}
\def \tr{{\textrm {Tr}}}

\begin{document}
\title{Converting separable conditions to entanglement breaking conditions}

\author{Ryo Namiki}
\affiliation{Institute for Quantum Computing  and Department of Physics and Astronomy,
University of Waterloo, Waterloo, Ontario, N2L 3G1, Canada}

\date{\today}
\begin{abstract} 
We present a general method to derive 
entanglement breaking (EB) conditions for continuous-variable quantum gates. We start with an arbitrary entanglement witness, and reach an EB condition. The resultant EB condition is applicable not only for quantum channels but also for general quantum operations, namely,  trace-non-increasing class of completely positive maps.  We illustrate our method  associated with a quantum benchmark based on the input ensemble of  Gaussian distributed coherent states. We also  exploit our idea for channels acting on finite dimensional systems and present a Schmidt-number benchmark based on input states of two mutually unbiased bases and  measurements of generalized Pauli operators. %
 \end{abstract}

\maketitle



An important task for future realization of  quantum information technology is  to establish a reliable quantum channel. 
A powerful tool to estimate an experimental implementation of quantum gates is quantum process tomography. However, it is not always feasible to  measure the input-and-output relations for a set of tomographic complete states.  Instead of  tomographic approach, one may  be interested in probing a basic coherence of quantum channels using a small set of feasible input states.  The quantum benchmarks provide such a method based on the context of  quantum entanglement \cite{Pop94,Mass95,Horo99,Ham05,namiki07,Owari08,Cal09}. 
 A quantum benchmark  is typically determined by an upper bound of  an average fidelity achieved by a class of  quantum channels called entanglement breaking (EB) \cite{ruskai03a}.  If an experimental fidelity surpasses the fidelity bound  we can verify that any classical measure-and-prepare map  is unable to   simulate the channel.  Mathematically, it implies the Choi-Jamiolkowsky (CJ) 
 state of  the channel is entangled, hence, there exists, at least, one entangled input state whose  inseparability maintains under the channel action.  There have been several works to determine such classical fidelities \cite{Horo99,Fuc03,Ham05,namiki07,Namiki08,Namiki11} or other forms of EB limits \cite{Has09,Namiki-Azuma13x}. One can also apply the notion of EB limits to quantum operations, namely, trace-non-increasing class of  completely positive (CP) maps \cite{Yang14,Chir13}.  
In addition to  a proof of the inseparability in the physical process, one can  demonstrate a more specified type of channel's coherence by quantifying  the amount of entanglement in the CJ state \cite{Namiki12R,Kil12,Khan13}.

Although  it has been known that an EB condition is mathematically equivalent to a separable condition,  the varieties of known EB conditions are rather limited compared with those of known separable conditions.  In fact, one can easily find  several systematic methods to produce  a series of separable conditions  \cite{SV,Miran,Miran10} whereas potential applications of separable conditions to the  quantum benchmark problems have little been mentioned in the literatures on the separability problems \cite{Horo09,Guhne09}.

In this report, we present a general method to convert a separable condition to an EB condition for continuous-variable quantum channels as a generalization of the method developed in \cite{Namiki-Azuma13x}. Given a formula of entanglement witness we compose an EB condition by separately assigning an entangled density operator.  After a general composition we illustrate our method  associated with   a quantum benchmark based on the Gaussian distributed coherent states \cite{Chir13}.   
We also exploit our idea for channels acting on finite dimensional systems and present a Schmidt-number benchmark \cite{Namiki12R} for qudit channels based on input states of  two mutually unbiased bases and  measurements of generalized Pauli operators.

Let $\rho$ be a density operator and write an expectation value of an operator $\hat O $ as $\tr[ \hat O \rho ]= \ave{ \hat  O}_\rho$. 
A general form of separable conditions can be written by a function of expectation values for a set of operators $\{ \hat  O_i\}_{i= 1,2, \cdots, N}$ as  
\begin{eqnarray}
F(\ave{ \hat O_1}_\rho ,\ave{ \hat  O_2}_\rho, \cdots, \ave{\hat  O_N}_\rho )\ge 0. \label{gsepcon}
\end{eqnarray}
  A special case is based on an operator called the {\it entanglement witness} $\hat W $  that satisfies   
 \begin{eqnarray}
\tr (\hat W \rho_s) = \ave{\hat W}_{\rho_s} \ge 0  \label{defwit}
\end{eqnarray}
 for any separable state $\rho_s = \sum p_i (\ketbra{a_i}{a_i})_A\otimes (\ketbra{b_i}{b_i})_B$. 
 It implies that $\rho$ is entangled when $\langle{\hat W}\rangle _{\rho} < 0$ holds. In what follows we derive an EB condition starting from an witness operator $\hat W$. We can readily extend our method for  the general form in  Eq.~\eqref{gsepcon}. This form includes non-linear terms of expectation values and is often referred to as the {\it non-linear witness}.

We consider a two-mode system $AB$ described by bosonic field operators satisfying the  commutation relations $[a, a^\dagger]= [b, b^\dagger]=1 $. 
Let us suppose $\hat  W$ is expressed in the anti-normal order regarding to the field operators $\{b, b^\dagger\} $  for the second system $B$ such as
\begin{align}
\hat  W (a,a ^\dagger, b, b^\dagger)  = \sum_{n,m} W^{(n,m)}(a, a^\dagger) b^n (b^\dagger)^{m}.
\end{align}
Then, we can rewrite it as
\begin{align}
\hat  W= &\sum W_{n,m}(a, a^\dagger) b^n  \openone_B (b^\dagger)^{m}  \nonumber \\
  =&\sum   W^{(n,m)}(a, a^\dagger)  \int (\alpha ^*)^n \alpha^m  \ketbra{\alpha^*}{\alpha^*}\frac{d^2 \alpha }{\pi} \nonumber \\ 
 =&   \int  \hat   W (a, a^\dagger, \alpha^*, \alpha ) \ketbra{\alpha^*}{\alpha^*}\frac{d^2 \alpha }{\pi}, \label{ik4}
\end{align} where we used  the closure relation for coherent states $\int \ketbra{\alpha}{\alpha } {d^2 \alpha }/ {\pi} = \openone$ 
for the subsystem $B$. Here, we express the closure with $\alpha^*$, the complex conjugate of $\alpha $,  for a notation convention.  Equation~\eqref{ik4} implies 
\begin{align}
 \tr(\hat   W \rho ) = &\tr_A \left[ \int  \hat  W (a, a^\dagger, \alpha^*, \alpha ) \bra{\alpha^*}\rho \ket{\alpha^*}_B\frac{d^2 \alpha }{\pi} \right],  \label{eq5}
\end{align} where $\tr _A$ denotes partial trace over system $A$.

Let  $\psi = \psi_{AB} $ be an entangled density operator  of  the two-mode field $AB$.  We define an ensemble of states $\{p_\alpha, \varphi_ \alpha\}$ on  a one-mode system  
  as 
\begin{align}
p_\alpha   &: =  \tr\left[  \openone_A \otimes ( \ketbra{\alpha ^*}{\alpha ^*})_B    \psi_{AB}  \right] , 
\nonumber \\
\varphi_ \alpha &: =  \bra{\alpha ^* } \psi_{AB} \ket{\alpha^* }_B  /  p_\alpha.  \label{em1} 
\end{align}  
Note that $\varphi_\alpha$ is a type of the  relative states of $\ket{\alpha ^*}$  regarding $\psi_{AB} $ and $p_\alpha $ is a probability density satisfying $\int  p_\alpha d^2 \alpha/ \pi  =1 $.

 Let us consider  the local action of a physical map $\mathcal E$ for the state $\psi$, 
\begin{align}
J = {\cal E}_A\otimes I_B (\psi ) \label{JState}
\end{align} where 
$\mathcal E$ is  a CP map acting on system $A$ and $I$ is the identity map. 
When $\mathcal E$ is a trace-decreasing operation,  we can formally normalize $J$ as a density operator by $J / P_s$ with 
  \begin{align}
P_s   &:   =  \tr [J ]= \int p_\alpha \tr [ \mathcal E ( \varphi_ \alpha )]    { d^2 \alpha }/{\pi}  , \label{em2}
\end{align} where we use the relations in Eq.~\eqref{em1}. 
  Note that we have $P_s = 1  $ for the trace-preserving maps.  
Substituting $\rho = J/P_s$ into Eq. \eqref{eq5} we can write 
\begin{align}
 \tr(\hat  W \rho ) = &\frac{1} {P_s } \tr \left[ \int   { \hat   W (a, a^\dagger, \alpha^*, \alpha ) p_\alpha {\cal E} (\varphi_ \alpha )} 
  \frac{d^2 \alpha }{\pi} \right].  \label{eq10}
\end{align}
 Here, system $B$ is traced out and $ \tr( \hat W \rho )$ is represented by the mean values of  operators on system $A$ over  channel's outputs  $\mathcal E (\varphi_\alpha)$ subjected to the input state  $ \{\varphi_\alpha\}$.

Let us  suppose that $\cal E$ is an EB map, i.e., $\mathcal E(\rho ) = \sum_i \tr [ M_i  \rho ] \sigma_ i$ with $ M_i  \ge 0$,  $ \sum_i M_i \le \openone$, and a set of density operators $\{ \sigma_i\}  $. Then, $\rho$ becomes a separable density operator and $\tr (\hat  W \rho )$ has to fulfills 
the separable condition of Eq.~\eqref{defwit}. Therefore, we  obtain the following  EB condition:  
\begin{align}
  \frac{1} {P_s } \tr \left[ \int   { \hat   W (a, a^\dagger, \alpha^*, \alpha ) p_\alpha {\cal E} (\varphi_ \alpha  )} 
  \frac{d^2 \alpha }{\pi} \right] \label{eq11} \ge 0 . 
\end{align}
 In this manner one can compose an  EB condition from a separable condition by separately assigning an entangled state $\psi$.  To be concrete,  the inequality of Eq.~\eqref{eq11} is a necessary condition for entanglement breaking, and any violation of this inequality implies that the map $\mathcal E$ cannot be an EB map. 

For a non-linear witness  in the form of Eq.~\eqref{gsepcon}, we simply assign  an  operators $\hat  W_i $ for  each of  $\hat O_i$ and  express its expectation value as in Eq.~\eqref{eq10} by repeating the procedure above. 
Then, we can generally convert separable conditions in the form of Eq.~\eqref{gsepcon}  into  EB conditions by replacing the relevant expectation values as  follows:  
\begin{align}
\ave{\hat O_i}_ \rho \to   \frac{1} {P_s } \tr \left[ \int   { \hat    W_i (a, a^\dagger, \alpha^*, \alpha ) p_\alpha {\cal E} (\varphi_ \alpha  )} 
  \frac{d^2 \alpha }{\pi} \right]  .
\end{align}
Note that the obtained EB condition depends on the choice of the entanglement $\psi$ which determines the state ensemble $\{p_\alpha, \varphi_ \alpha   \}$ owing to Eq.~\eqref{em1}. 
Accordingly, a different choice of $\psi$ could  constitute a different EB condition  even the original separable condition is the same.

Let us illustrate our method associated with a familiar case of the fidelity-based quantum benchmark \cite{Ham05,namiki07,Namiki11,Chir13}. 
In experiments of quantum optics,  coherent states are commonly  available  as a state of laser light. It is thus feasible to probe an experimental quantum gate by an input of coherent states.  We will consider  
an input ensemble of  the Gaussian distributed coherent states 
\cite{Bra00}.  This ensemble can be associated with the case that $\psi$ is a two-mode squeezed state.    
In fact, by substituting  the two-mode squeezed state $\ket{\psi_\xi}= \sqrt{1-\xi ^2} \sum_{n=0}^\infty \xi^n\ket{n}\ket{n}$ with   $\xi \in (0,1)$ into Eqs.~\eqref{em1}, 
 we obtain the ensemble of Gaussian distributed coherent states,  
\begin{align}
p_\alpha   & =  (1- \xi^2) e^{-(1-\xi^2)|\alpha |^2 }  , \nonumber \\
 \varphi_ \alpha &  =   \ket{\xi \alpha  }\bra{\xi \alpha  } \label{GDC} . 
\end{align}

Let  $X \ge 0 $ and $(u,v)$ be a pair of real number that fulfills $u^2 + v^2 =1  $ and $u \neq 0$. Let us define  an witness operator
\begin{align}
\hat W :  =&  \frac{\openone}{ 1+X} - \frac{1}{\pi}  \int e^{-X |\alpha|^2} \ketbra{v \alpha}{v \alpha}  \otimes\ketbra{u\alpha^*}{u\alpha^*}   { d^2 \alpha }  , \label{eq14}
\end{align}  such that $\langle \hat W \rangle\ge 0$ becomes  the separable condition in Eq.~(21) of Ref.~\cite{Namiki13J}.
   Since $\hat W $ is already expanded in the local coherent states similar to the form in Eq.~\eqref{ik4} it is  no need to consider the operator ordering.   
From Eqs.~\eqref{eq10},~\eqref{GDC},~and~\eqref{eq14} we can write 
\begin{align} 
 \tr [  \hat W J ]=&  \frac{1}{ 1+X} - \frac{1}{\pi P_s  u^2 } \left( \lambda + \frac{X}{\xi^2 u ^2 }\right)   \nonumber \\ &  \times \int e^{-\lambda  |\alpha|^2} \bra{ \sqrt \eta  \alpha}\mathcal E (\ketbra{ \alpha }{ \alpha }) \ket{ \sqrt \eta  \alpha}  { d^2 \alpha }   , %
\end{align}  
where $\lambda = \xi^{-2}( X u^{-2}+ (1- \xi^2)) $ and $\eta:= v^2 / (\xi u )^2$.  Using the condition of Eq.~\eqref{eq11} and taking the limit $X \to 0 $ we obtain the following  EB condition
\begin{align} 
  &   {P_s }  - \frac{1 }{u^2 } \frac{\lambda }{\pi} \int e^{-\lambda  |\alpha|^2} \bra{ \sqrt \eta  \alpha}\mathcal E (\ketbra{ \alpha }{ \alpha }) \ket{ \sqrt \eta  \alpha}  { d^2 \alpha }   \ge 0,  
\end{align}   where $u^2 =  (1+ \lambda + \eta )/(1+ \lambda )$. This corresponds to the fidelity-based quantum benchmark for general CP maps \cite{Chir13}.  In Ref.~\cite{Chir13},  its derivation is based on the duality of semidefinite programing. 
 For quantum channels (the trace-preserving class of CP  maps; $P_s =1$), one can find other derivations in Refs.~\cite{Ham05,namiki07,Namiki11}.


Note that there is a wide interest in formulating  separable conditions based on the moments of  canonical quadrature variables \cite{Duan,Simon,Giov03,SV}. The moments of quadrature variables can be directly observed by homodyne measurements in experiments. Among all, second-order conditions have been  widely used as a feasible method for entanglement detection.  It is well-known that  the sum condition \cite{Duan} and the product condition \cite{Giov03}  are sufficient  for witnessing two-mode Gaussian entanglement. 
By applying our method we can translate them into the EB conditions with the input ensemble of the Gaussian distributed coherent states in \cite{Namiki-Azuma13x}  (Corollary 1 and Proposition, respectively), which are sufficient to witness one-mode Gaussian channels in the quantum domain, namely,  one-mode Gaussian channels being nonmember of  the EB class. Further, the formalism developed in Ref.~\cite{Namiki-Azuma13x} would be usable as  a quantitative quantum benchmark because it can be related to entanglement of formation on the CJ state (See Ref.~\cite{Namiki15}).  Similar statements could hold for the fidelity-based approach. In fact,  the entanglement witness of Eq.~\eqref{eq14} is known to be sufficient  for witnessing two-mode Gaussian entanglement \cite{Namiki13J} and  the fidelity-based EB condition is also sufficient for detecting  one-mode Gaussian channels in the quantum domain \cite{namiki07}. However, its connection to a meaningful entanglement measure   is left open. 

In the rest of this report, we discuss the case of the physical process acting of a finite dimensional system. The  key mechanism to introduce  the ensemble of input states $\{p_\alpha, \varphi _\alpha\}$  in Eq.~\eqref{em1}  
  is the coherent-state expression of  system $B$ in  Eq.~\eqref{ik4}. Analogously, we can introduce a state ensemble   by  decomposing the witness operator  with a set of hermitian operators $\hat h$ on system $B$ as follows
\begin{align} 
\hat W &= \sum_{l} w_A^{(l)} \otimes {\hat h}_B ^{(l)} = \sum_{l} w_A^{(l)} \otimes  \left( \sum_j  { h_j} ^{(l)} |j^{(l)}\rangle  \langle j^{(l)}| 
 \right)_B,    \label{fideco}
\end{align}  where $\{ h_j^{(l)}, \ket{j^{(l)}}\}$ represents the spectral decomposition of $\hat h^{(l)} $. This implies 
 the set of input states similarly to Eq.~\eqref{em1} as 
  \begin{align}
p_{j,l }   &: =  \tr\left[  \openone_A \otimes ( |j^{(l)}\rangle \langle  j^{(l)}| )_B    \psi_{AB}  \right] 
\nonumber \\
\varphi_{j,l}  &: =  \langle  j^{(l)}| \psi_{AB} |j^{(l)}\rangle _B  /  p_{j,l}.  \label{em1-} 
\end{align}  Therefore, instead of Eq.~\eqref{eq11}, we obtain an EB condition  in the following form:  
\begin{align}
  \frac{1} {P_s } \sum_{j,l } p_{i,l}  h_j^{(l)}  \tr \left[    {   \hat w ^{(l)}{\cal E} (\varphi_{j,l } )}  \right]  \ge 0,  \label{witten}
\end{align} where we define $P_s =  \sum_{j,l } \tr [ p_{j,l} {\cal E} (\varphi_{j,l } )    ] $. 
Note that an example of the decomposition in Eq.~\eqref{fideco} can be obtained  by choosing  a Hilbert-Schmidt orthonormal  basis on subsystem $B$. %

Finally, using this framework we will derive a Schmidt number benchmark \cite{Namiki12R} for quantum operations acting on a $d$-dimensional (qudit)  system.    
The Schmidt number benchmark of class $k+1$ ($k \in [1,d-1]$) enables us to eliminate the possibility that the channel is described by Kraus operators of rank $k$ or less than $k$. This class of quantum channels is called  $k$-partial EB channels \cite{Chru06,Hua06,Namiki13a}, and $k=1$ represents the class of EB channels.

Let us consider a Schmidt number-$(k+1)$ witnesse for two $d$-dimension system given in  Ref.~\cite{Namiki12L}, 
 \begin{eqnarray}
g_{k,d}  - \frac{1}{2}\ave{\hat Z_A\hat Z_B^\dagger + \hat Z_A^\dagger \hat  Z_B +\hat  X_A\hat  X_B +\hat X_A ^\dagger \hat   X_B ^\dagger} \ge 0 ,  \label{ex2}
\end{eqnarray} where $g_{k,d} =  [{(d-k ) \cos \omega + (d +k ) }]/{d }$,  and  
$\hat Z: = \sum_{j= 0}^{d-1} e^{ i\omega j} |j  \rangle\langle j|$ and 
$ \hat X: = \sum_{j=0}^{d-1}|j+1 \rangle\langle j|$,   are the generalized Pauli  operators. Here, we
 assumed a fixed $Z$-basis $\{ \ket{0},\ket{1}, \cdots, \ket{d-1}\}$ with modulo-$d$ conditions $|j+ d\rangle = |j\rangle $ and $\omega  := 2 \pi /d $. 
By expanding   $\hat Z$ and $\hat X$ respectively in $Z$-basis $\{\ket{j} \}$ and $X$-basis  $\{\ket{\bar  j} \}$, which is defined through  $| \overline {l } \rangle :=  \hat Z ^ l  \left(
\frac{1}{\sqrt d} \sum_{j=0}^{d-1 }\ket{j } \right) =\hat Z ^ l |\overline {0} \rangle$,   we can see that the operators on system $B$ in Eq.~\eqref{ex2}  can be expressed by the projections onto the mutually unbiased bases, $ \{ \ket{j}, \ket{\bar j } \}$. Using this expansion and  $J = \mathcal E_A \otimes I_B ( \ketbra{\Phi_d}{\Phi_d}) /P_s$ with $\ket{\Phi_d} = d ^{-1/2} \sum_{j=0}^{d-1} \ket{j}\ket{j}$  with   Eqs.~\eqref{em1-} and \eqref{witten}  we obtain 
 the following necessary condition for $k$-partial EB maps: 
\begin{align}
& P_s g_{k,d}  -  \sum_{j=0}^{d-1} \tr \big [(  \hat Z e^{-i \omega j  } + \hat Z ^\dagger  e^{ i \omega j }) \mathcal E ( \ket{j}\bra{j} ) \nonumber \\ &+   (  \hat X e^{-i \omega j  } + \hat X ^\dagger  e^{ i \omega j }) \mathcal E ( \ket{\overline {-j}}\bra{\overline {- j}} ) \big] /d   \ge 0 .   \label{ex2x}
\end{align} Hence, a   violation of this condition implies a  quantitative quantum  benchmark for the Schmidt class $k+1$, namely, any Kraus representation of $\mathcal E$ has,  at least, one Kraus operator whose rank is $k+1$ or higher. 
 An  experimental test would be executed by input states of two mutually unbiased bases and projections to these bases similarly to the result in Ref.~\cite{Namiki12R}. 
Note that  we can readily extend  the result in Ref.~\cite{Namiki12R}
 for quantum operations acting on qudit states by using the normalized state $J = \mathcal E_A \otimes I_B ( \ketbra{\Phi_d}{\Phi_d}) /P_s$.

In summary,  we have presented a method to  convert separable conditions to  EB conditions for bosonic single-mode channels. Given an entanglement witness we can generate an EB condition by  separately assigning an entangled state that determines the ensemble of  input states. 
 By considering a  normalization  of this state the resultant EB condition becomes applicable to  general quantum operations, namely, trace-non-increasing class of CP maps.  As an example we present a different derivation of the fidelity-based quantum benchmark in Ref.~\cite{Chir13} starting from a separable condition given in Ref.~\cite{Namiki13J}. 
  Although we focus on single-mode operations,  our method can be  straightforwardly extended for multi-mode bosonic quantum channels/operations.
 We have also developed  a similar framework for quantum operations acting on finite dimension systems and presented a Schmidt number benchmark for quantum operations.

\acknowledgments
RN was supported by the DARPA Quiness program under prime Contract No. W31P4Q-12-1-0017 and Industry Canada. 

\end{document}